# Stress, Strain, or Energy: Which One Is the Superior Parameter to Estimate Fatigue Life of Notched Components? An Answer by a Novel Machine Learning-Based Framework

A. M. Mirzaei[1]

*Department of Structural, Geotechnical and Building Engineering, Politecnico di Torino, Corso Duca degli Abruzzi 24, 10129 Torino, Italy*

**Abstract**

This paper introduces a simple framework for accurately predicting the fatigue lifetime of notched components by employing various machine learning algorithms applied to a wide range of materials, loading conditions, notch geometries, and fatigue lives. Traditional approaches for this task have relied on empirical relationships involving one of the mechanical properties, such as stress, strain, or energy. This study goes further by exploring which mechanical property serves as a better measure. The key idea of the framework is to use the gradient of the mechanical properties (stress, strain, and energy) to distinguish between different notch geometries. To demonstrate the accuracy and broad applicability of the framework, it is initially validated using isotropic materials, subsequently applied to samples produced through additive manufacturing techniques, and ultimately tested on carbon fiber laminated composites. The research demonstrates that the gradient of all three measures can be effectively employed to estimate fatigue lifetime, with stress-based predictions exhibiting the highest accuracy. Among the machine learning algorithms investigated, Gradient Boosting and Random Forest yield the most successful results. A noteworthy finding is the significant improvement in prediction accuracy achieved by incorporating new data generated based on the Basquin equation.

---

[1]Corresponding author. Tel.: +39 011 090 4910; Fax: +39 011 090 4899.

E-mail address: amir.mirzaei@polito.it

# 1. Introduction

Estimating fatigue life is a vital task in the design and maintenance of various mechanical systems and structures. This process involves predicting how many cycles (life) a material can endure before failure due to cyclic loading, a phenomenon commonly encountered in real-world structures. Traditionally, fatigue life estimation has relied on empirical relationships derived from extensive experimental testing, focusing on mechanical properties like stress, strain, or energy. Therefore, these methods can be time-consuming, expensive, and may lack precision for diverse materials or loading conditions. Recently, machine learning (ML) has emerged as a leading solution for fatigue life estimation. It offers the potential to enhance prediction accuracy by identifying patterns within existing datasets, revolutionizing the field. This section provides a review of the current state of the art in fatigue analysis through machine learning, initially exploring its broad applications across various problems and materials, then narrowing down to its use in additively manufactured samples, and ultimately presenting advanced models to set the context for the problem statement.

ML techniques have significantly improved the predictive accuracy of fatigue life estimation of various materials, as evidenced by several compelling applications. For instance, hybrid physics-informed and data-driven models (HPDM) have proven to be effective, laying a methodological foundation further substantiated by the successful application of the XGBoost model in predicting the fatigue life of high-strength bolts [1]. This approach has also been extended to the prediction of low-cycle fatigue life for lead-free solders, demonstrating ML's reliability across various fatigue conditions [2]. Another illustrative example is the development of a knowledge-based ML framework that integrates empirical formulas with data-driven models, resulting in improved accuracy and resource efficiency [3]. The versatility of ML is further exemplified by its successful application in predicting fatigue life in drilled composite laminates [4]. To emphasize the extensive applicability of machine learning techniques across various materials in the domain of fatigue life estimation, the following can be cited: A588 low alloy steel [5], 316 stainless steel [6], gray cast iron [7], cast aluminum alloy [8], multi-principal element alloys [9], and additively manufactured (AM) parts [10,11]. Moving to multiaxial loading, in a novel twist, image recognition technology has been utilized to extract loading path features for multiaxial fatigue life prediction [12].

Moreover, the fusion of virtual synthetic multiaxial fatigue data with ML models has further enhanced the accuracy of life prediction models, as proposed in [13].

Focusing on AM materials, ML techniques have been effectively employed to tackle the intricate nature of fatigue phenomena inherent to these materials. These methods have consistently proven to be indispensable in this context [14]. The effectiveness of ML in handling uncertainties and optimizing manufacturing processes to enhance fatigue performance is highlighted in studies such as [15–17]. To address the issue of limited fatigue data and to expand the dataset size for Ti-6Al-4V, Monte Carlo simulation in conjunction with ML models yielded promising results [18]. In noteworthy research [19], ML was combined with natural language processing and computer vision to collect and analyze fatigue and mechanical data from thousands of scientific articles. This resulted in the creation of a comprehensive database, FatigueData-AM2022, which can serve as a benchmark and training set for further ML-driven analyses. The application of machine learning has also extended to predicting microstructural-level fatigue damage [20] and fatigue behavior in various materials and contexts within AM, including laser powder bed fusion materials [6] and non-metallic materials like polymer composites [21].

Advancements in computational frameworks, such as the one proposed in [22], underscore the potential of ML in optimizing fatigue performance. A noteworthy development in this direction is represented in [23], where a Physics-Informed Neural Network was utilized to predict finite fatigue life in defective materials, showcasing the successful integration of traditional physics models with advanced ML techniques. Additionally, in a recent study [24], an enhanced offline prediction approach was introduced for predicting the lifespan of fatigue crack propagation. This method relies on data feedback and highlights the significance of iterative learning and adaptation within machine learning frameworks. It is worthwhile to mention that the emergence of ML techniques has brought about a transformative shift in fatigue life prediction strategies, leveraging innovative computational approaches. For instance, a hybrid model integrating fracture mechanics and ML has excelled in predicting high-cycle fatigue life [25]. A unified physics-informed ML framework has shown significant promise in predicting notch fatigue life, highlighting the effectiveness of incorporating physical principles into ML algorithms [26]. These algorithms, including the application of shallow neural networks and sequence learning models, were utilized to forecast the lifetime under isothermal low-cycle fatigue loading and thermo-mechanical fatigue loading, yielding accurate estimations for these complex loadings [27].

In summary, the integration of machine learning techniques with traditional methods has revolutionized fatigue life estimation, offering enhanced accuracy, adaptability, and efficiency across various materials and loading conditions. This research aims to build upon these advancements, exploring the specific application of ML in fatigue lifetime assessment of notched structures. In the following section, a comprehensive introduction to the problem and the objective of the research is provided.

## 2. Problem statement

Geometric discontinuities, such as cracks, notches, and holes, can be found within mechanical structures, serving specific design intentions or arising from the effects of wear and environmental factors. These elements, acting as stress concentrators, play a pivotal role in delineating both the load-bearing capability and operational lifespan of the structural component. However, due to its complexity, this topic is less often addressed in the literature. In this paper, a novel framework is proposed to estimate the fatigue life of notched components using ML algorithms, based on various mechanical parameters such as stress, strain, or energy fields.

The inspiration for this work stems from Neuber's study [28], which introduced a successful approach for fatigue analysis of notched components. This approach later became known as the Theory of Critical Distances [29,30] and has recently been incorporated into the Finite Fracture Mechanics approach [31–33]. These strategies typically depend on the *stress field* around the notch tip to distinguish various notch geometries. Fig. 1 provides a schematic illustration of the stress field corresponding to two distinct notch geometries, indicating that a sharper notch corresponds to a higher stress gradient [34].

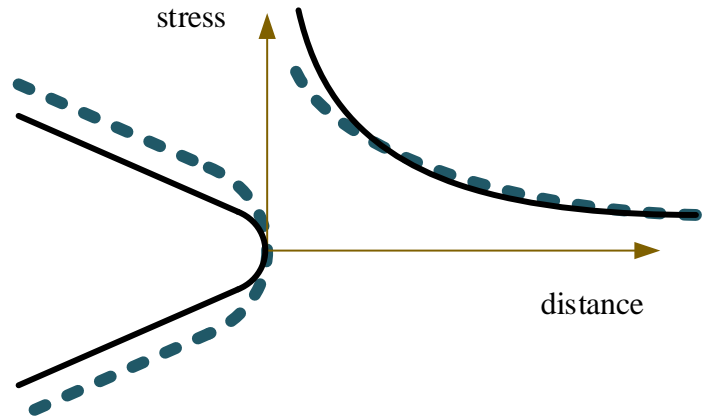

Fig. 1. Schematic illustration of two different notch geometries with their related stress distribution.

On the other hand, studies have shown that certain materials are subject to strain-controlled failure [35]. As a consequence, the strain field is incorporated into our analysis. Finally, the energy release rate, widely used in Linear Elastic Fracture Mechanics (LEFM) and its applications, is utilized in this study for the sake of comparison and completeness. It is worthwhile to mention that linear elastic analysis is employed to derive stress, strain, and energy release rate fields. This methodology is especially pertinent in scenarios involving medium- and high-cycle fatigue loading conditions, where even ductile materials typically exhibit linear elastic behavior.

The primary aim of this research is to propose an innovative framework that incorporates either stress, strain, or energy fields along with machine learning algorithms to estimate the fatigue lifetime of notched components. Moreover, this study seeks to identify the most effective field for this purpose. In the next section, the ML tools that are employed for establishing the script are presented, while the measures and inputs of the model are discussed in Section 4.

## 3. Machine learning tools

This section provides an introduction to the script and the machine learning algorithms that have been utilized within this study.

Initially, the script emphasizes data preprocessing. If activated, useful in cases where the original dataset is limited in size, new data can be synthetically generated using a function that resembles a line in log-log space, with an added noise factor ranging between -0.3 and 0.3, as shown in Fig. 2. In this scenario, the code ensures that the amount of data for each notch geometry is set to 100 (will be elaborated in Section 5.5.4). This technique for generating synthetic data mimics real-world scenarios where data often exhibit a power-law distribution, causing the data to form a straight line in log-log space according to the Basquin equation [36]. The noise introduced helps emulate the unpredictable variations typically found in real data, providing a more realistic testing ground for the regression models. This approach offers valuable insights into model performance under varying and potentially noisy conditions. In the analyses, the term *baseline data* refers to the original data, while *enhanced data* refers to the combination of the original and generated data.

To capture complex relationships within the data, the script initially generates new features by summing up the product of applied stress and all other columns (will be explained in Section 4). Subsequently, it handles categorical data (a name related to each notch shape) by converting them into numerical format through one-hot encoding. The script also computes correlations between features and the target variable, applying a feature selection strategy based on a predetermined threshold to retain only the most relevant features. Section 4 discusses the features specific to each problem. Moreover, the script offers flexibility in handling unseen data by either randomly selecting from the dataset or using specific types of data. Unseen data play a crucial role in evaluating the performance of machine learning models on untrained data. To enhance model performance, the data are normalized using RobustScaler, which effectively handles outliers. Normalizing the data ensures that all features are on a consistent scale, preventing dominance by features with larger scales and promoting balanced model performance.

After the data preprocessing, the script employs five regression models for training and validation, utilizing K-fold cross-validation with five folds. The models used encompass Gradient Boosting, Random Forest, Linear Regression, Multi-Layer Perceptron, and a Voting Regressor that amalgamates the aforementioned models. For simplicity, these models are trained with their default settings, without any customization or tuning of their parameters, thereby facilitating the model coding process. Cross-validation is employed to assess the models' ability to generalize to unseen data. The code executes 100 iterations for each regression model, enabling a more comprehensive performance evaluation and accounting for variability in data selection.

In selecting the specific regression models for this study, a diverse and effective approach was considered to predict the fatigue lifetime of notched components. Gradient Boosting and Random Forest were chosen for their proficiency in handling complex non-linear patterns and their consistent success in the study's outcomes. Linear Regression serves as an easily interpretable baseline model, and the Multi-Layer Perceptron is utilized for its ability to explore intricate non-linear relationships. The Voting Regressor uniquely integrates the strengths of all the aforementioned models, providing an additional layer of robustness. Furthermore, an essential consideration in the selection of these models was their relative ease of coding and implementation compared to other more complex machine learning approaches. This decision not only facilitated the model coding process but also ensured a practical and accessible methodology. By combining simplicity in coding with both linear and complex modeling capabilities, this ensemble offers a well-balanced, nuanced, and efficient approach to the prediction task, aligning seamlessly with the multifaceted nature of the variables and conditions investigated.

Having previously described the workflow of the prediction procedure and its constituent components, we now turn our attention to the specific models employed in our study. Concise introductions to each of these models are provided to offer a more in-depth reasoning behind choosing the models and their predictive ability for the problem, thus setting the stage for a more comprehensive discussion on the operation and performance of the predictive framework.

### 3.1 Gradient Boosting Regressor

Gradient boosting is a machine learning technique where weak predictive models, typically decision trees, are combined to form a powerful ensemble model [37,38]. It is a stage-wise optimization algorithm that can accommodate a variety of loss functions [39]. The principle of gradient boosting involves iteratively building weak learners and adding them to the ensemble model in such a way that the errors of the previous models are minimized. This is achieved by fitting each new model to the residuals or errors of the preceding models:

$$e_i = y_i - F_{m-1}(x_i) \tag{1}$$

where $e_i$ is the residual for the *i*-th instance, $y_i$ is the actual output for the *i*-th instance, $F_{m-1}(x_i)$ is the prediction of the (*m-1*)-th model for the *i*-th instance. A weak learner is then fit to these residuals. This results in a new model, represented as:

$$F_m(x_i) = F_{m-1}(x_i) + h_m(x_i) \tag{2}$$

where $h_m(x_i)$ is the prediction of the *m*-th weak learner for the *i*-th instance.

This iterative process results in an ensemble model, composed of a sum of weak learners:

$$F(x_i) = F_0(x_i) + \sum_{m=1}^{M} h_m(x_i) \tag{3}$$

where $F_0(x_i)$ is the initial model's prediction for the $i$-th instance $M$ is the total number of weak learners in the model [40].

## 3.2 Random Forest Regressor

Random Forest (RF) is a machine learning approach where multiple decision trees, each with unique characteristics, are combined to form a robust ensemble model [41,42]. The fundamental principle of RF lies in constructing a multitude of decision trees and aggregating their predictions to determine the final output. The construction of each tree, denoted by $T_m(x)$, is based on a bootstrap sample drawn from the training set with replacement. Further, at each node, a random subset of $m$ variables is chosen out of the total $p$ variables, and the optimal split among these $m$ variables is used to bifurcate the node [43].

The prediction of the RF model for an instance is given by averaging the predictions from all the decision trees. This can be mathematically represented as:

$$F(x_i) = \frac{1}{N} \sum_{m=1}^{N} T_m(x_i) \tag{4}$$

In this equation, $F(x_i)$ is the RF prediction for the $i$-th instance, $N$ represents the total number of trees in the RF model, and $T_m(x_i)$ denotes the prediction of the $m$-th tree for the $i$-th instance.

Each decision tree independently draws its bootstrap samples from the dataset:

$$D_m = \text{bootstrap}(D), \quad \text{for } m = 1 \text{ to } N \tag{5}$$

where $D_m$ is the dataset for the $m$-th tree, $D$ signifies the original dataset.

Despite the many advantages of RF, it is crucial to remember that, unlike gradient boosting which builds models in a stage-wise additive manner, decision trees in an RF model are constructed and combined independently.

## 3.3 Linear Regression

Linear regression is a statistical technique widely used in machine learning to predict a dependent variable based on the values of one or several independent variables [44]. This approach, foundational in statistics and machine learning alike, adopts a systematic method for determining the relationship between predictors and outcomes [45]. The principle of linear regression involves fitting a model to the data in such a way that the sum of squared residuals (SSR) is minimized. The residual for an instance is calculated as the difference between the actual output and the prediction of the model:

$$e_i = y_i - \left(\beta_0 + \sum_{j=1}^{n} \beta_j X_{ij}\right) \qquad (6)$$

where $e_i$ is the residual for the $i$-th instance, $y_i$ is the actual output for the $i$-th instance, $\beta_0$ is the y-intercept, $\beta_j$ is the coefficient of the $j$-th independent variable, $X_{ij}$ is the value of the $j$-th independent variable for the $i$-th instance. The model aims to determine the coefficients ($\beta_j$) that minimize the sum of squared residuals. This is achieved through techniques such as ordinary least squares (OLS) or gradient descent for larger datasets due to its computational efficiency [46].

The final prediction of the model is given by the equation:

$$Y_i = \beta_0 + \sum_{j=1}^{n} \beta_j X_{ij} \qquad (7)$$

where $Y_i$ is the predicted output for the $i$-th instance.

While the simplicity and interpretability of linear regression models have made them a mainstay in many scientific disciplines, it is important to bear in mind that these models rely on certain assumptions. These include linearity, independence of errors, homoscedasticity, and normality of errors. Violations of these assumptions can lead to model inaccuracies or misleading results.

### 3.4 Multi-Layer Perceptron (MLP) Regressor

The Multilayer Perceptron (MLP) is an artificial neural network featuring multiple interconnected layers, each serving a unique role in the model's architecture [47,48]. MLP uses non-linear activation functions and leverages backpropagation for iterative weight adjustment, which minimizes the error between the actual and predicted output.

The process of weight adjustment is expressed as:

$$e_i = y_i - F_{n-1}(x_i) \qquad (8)$$

where $e_i$ represents the error for the $i$-th instance, $y_i$ is the actual output, and $F_{n-1}(x_i)$ is the previous model's prediction for the $i$-th instance. The weights of the MLP are then updated:

$$F_n(x_i) = F_{n-1}(x_i) + h_n(x_i) \tag{9}$$

where $h_n(x_i)$ denotes the change in the weight for the *i*-th instance in the *n*-th iteration.

The iterative process culminates in an MLP model, which is the assembly of layered nodes:

$$F(x_i) = F_0(x_i) + \sum_{n=1}^{N} h_n(x_i) \tag{10}$$

where $F_0(x_i)$ is the initial prediction for the *i*-th instance and $N$ is the total number of iterations.

MLPs are used in various applications due to their ability to map input features to continuous output variables. Unlike gradient boosting models, the MLP's layers are interconnected and weights are adjusted simultaneously, enabling MLPs to capture complex non-linear relationships in the data.

### 3.5 Voting Regressor

The Voting Regressor is a machine learning technique that utilizes multiple base regressors, each trained on the entire dataset, to form an aggregate model by averaging predictions. This ensemble method can include a diverse range of algorithms such as Decision Trees, Support Vector Machines, or Neural Networks as its base regressors [49,50]. The process of averaging helps to reduce variance when the base regressors are prone to overfitting the data [49] and also to decrease bias when they are inclined to underfit the data [51].

Considering its limitations, the standard practice of utilizing uniform weights for ensemble voting regression may prove to be insufficient for improving predictions without the aid of domain knowledge. The performance of the regressor can also be impacted by the different voting methods used, and in some instances, more computationally expensive methods like the single transferable vote may be necessary.

### 4. Inputs and data for validation

To conduct a thorough evaluation of the proposed framework's performance and reliability in estimating finite fatigue life of notched components, it is essential to work with robust and extensive datasets. This requirement becomes even more critical when determining the most effective parameter between stress, strain, and energy. To achieve this objective, three distinct sets of experimental data were incorporated into the study, encompassing a range of notch geometries, including U-notch, V-notch, and circular holes. They also cover different materials such as steel, Poly Lactic Acid (PLA), and long carbon fiber laminated composites, all subjected to different (uniaxial) loading conditions and ratios. Each dataset possesses unique characteristics, providing a diverse and comprehensive foundation for our evaluation. This approach allows us to maintain a balance between complexity and comprehensiveness, ensuring that our analysis of the proposed framework's performance remains insightful and concise. Note that the stress-life (SN) diagrams for each dataset are not presented for the sake of brevity. Readers interested in further details are referred to [29,52,53].

In order to calculate the stress and strain distributions, along with the energy release rate, the finite element (FE) method was used. Given that this paper focuses on addressing uniaxial (mode I) loading conditions, the assumption is made that the failure happens in a direction perpendicular to the loading condition (a more comprehensive explanation of this assumption will be provided in this section). Hence, within the scope of the analyses, the terms stress, strain, and energy fields refer to the maximum principal (hoop) stress component along the notch bisector line, the maximum principal strain component along the notch bisector line, and the energy release rate pertaining to a crack situated at the notch tip and growing along the notch bisector line. The samples were discretized using 8-node biquadratic elements, and a convergence analysis was performed to verify the independence of the numerical results from the mesh size. Then, the stress, strain, and energy fields were extracted up to a sufficient distance where, for example, the stress field reached a (near-) constant value (it is discussed in Section 5.5.3). A total of 10 data points were utilized, with a higher density (gradient) of points near the notch tip (it is discussed in Section 5.5.1). The specimens were simulated under remote stress with amplitude of 1 (MPa). Therefore, by multiplying the gross stress amplitude from stress-life (SN) data by the stress, strain, or energy fields obtained through FE analyses, the corresponding fields around the notches related to each loading condition are derived. Note that these data, along with the corresponding number of cycles to failure (life) and a name for each notch shape (which is transformed into a numerical form through one-hot encoding), constitute the only input vectors for the problem. In other words, the inputs include the stress-life data, a name of the notch shape as well as stress, strain, or energy fields for a stress amplitude of 1 (MPa). As mentioned in Section 2, each of the stress, strain, or energy fields is unique to a specific notch geometry, and these fields make distinctions between different geometries. All in all, it can be argued that the proposed framework is simple, as it requires only linear elastic analyses for each notch geometry. On the other hand, in terms of machine learning coding, simple algorithms with the default settings without tuning of their parameters are employed. The subsequent section will provide a more detailed discussion of this process, as well as highlighting the key attributes of each dataset, and their distinctive characteristics and significance within the evaluation.

**4.1 Samples made of steel**

The first data set comprises samples made of EN3B steel and encompasses both plain and notched specimens, with six distinct notched geometries, each tested under tension-tension loading ($R = 0.1$) [29]. These tests were conducted under either tensile or bending loading conditions. For the specimens under tensile loading, the samples were weakened by a single edge V-notch with an opening angle ($\alpha$) of 60° and a notch tip radius of 0.12 mm, a single edge U-notch ($\alpha = 0°$) with a radius of 1.5 mm, and two central circular holes with two different diameter of 3.5 mm and 8 mm. Specimens under bending loading condition, which were approximately four times thicker than the tensile samples, contained a single edge V-notch with an opening angle of 45° and a radius of 0.383 mm, as well as a single edge U-notch with a radius of 5 mm.

To illustrate the performance of the code for generating new data, Fig. 2 presents the stress-life diagrams for the plain and V-notch samples under tension.

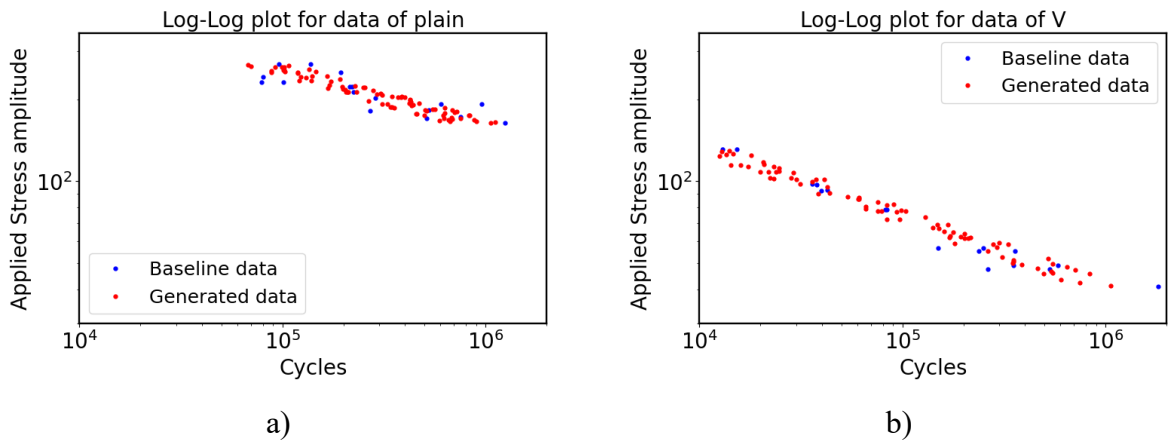

Fig. 2. Stress-life diagrams illustrating the baseline and enhanced (baseline+generated) data for two distinct specimens made of steel: a) Plain specimen, and b) V-notched specimen under tension.

As mentioned, if the option of increasing the number of data is activated, it is ensured that the total number of data points (SN data) for each type of notch geometry in enhanced dataset is equal to 100, following the Basquin equation with some noises.

### 4.2 Samples made of PLA

To further evaluate the performance of the proposed framework for finite fatigue life estimation, the experimental dataset presented in [52] is considered. These experiments involved testing PLA samples with three different manufacturing angles ($\theta_p$ = 0°, 30°, 45°) under loading ratio of $R$ = -1. Apart from the plain samples, the specimens were configured as double-edge notched tension, with a V-notch with an opening angle of 35° and a notch tip radius of 0.15 mm, as well as two U-notches with radii of 1 mm and 3 mm. An extensive experiment led the authors to argue that the material could be modeled as linear homogeneous and isotropic by averaging the mechanical properties across the different manufacturing angles [54]. Consequently, the effects of shell thickness and material non-linearities were considered negligible. Thus, the stress-strain behavior was modeled as purely linear elastic, up to the ultimate tensile strength.

### 4.3 Samples made of laminated carbon fiber composites

The last dataset consists of specimens made of T300-12K carbon fiber and Araldite LY 5052 epoxy resin, which were cured using Aradur 5052 Hardener [53]. The testing campaign included a plain sample, notched samples - center holed with diameter of 4.2 mm and 6.5 mm - as well as a center cracked sample with length of 15.4 mm. These samples were tested under tension-tension loading with a ratio of $R$ = 0.1 and had [0/90]$_{2s}$ layup. It is important to note that, due to the use of thin-ply laminate configuration, the specimens experienced failure in either the pull-out or brittle failure modes, without any delamination occurring and the failure was along the notch bisector line (perpendicular to the loading direction). This characteristic enables the use of stress, strain, and energy fields along the notch bisector line for failure analysis and estimation [55].

### 5. Results and discussion

This section presents the results of the proposed framework applied to various experimental datasets. It also includes a comprehensive examination through several parametric studies to evaluate the impact of different choices on data selection. The predictions for stress, strain, and energy are showcased for both enhanced and baseline data across each dataset. For each dataset, detailed tables are also presented to illustrate the performance of all the ML models.

In order to provide a robust estimation of the model's performance, the results include the average values of the $R^2$ score and mean squared error (MSE) obtained from 100 different episodes or iterations of the code. Importantly, each episode employs a unique split of training and validation data, generated by a 5-fold cross-validation scheme. This means that in each episode, the model is trained on 80% of the data and validated on the remaining 20%, with the specific data points allocated to each subset changing in every episode. Furthermore, the best $R^2$ score achieved among all model runs, whether for baseline or enhanced data, is highlighted, giving valuable insights into the model's peak performance capability. This strategy of averaging performance metrics across multiple episodes and observing the best score not only gives a more comprehensive and reliable understanding of the model's overall performance but also provides a sense of the model's potential under optimal conditions.

For all three datasets, an initial 10% of the data is considered as unseen data, and predictions are generated for this subset. These predictions are depicted in Figs. 3-5, where panels a) to c) correspond to cases with enhanced data (shown in Fig. 2), and panels e) to g) represent the baseline data (without any new data production). In all related figures, the horizontal axis represents the estimated life (Predictions), while the vertical axis illustrates the experimental number of cycles (Original Data). The solid black line represents a perfect estimation with zero error. The region above the solid line corresponds to conservative predictions, while the region below indicates unconservative predictions. The dashed red lines represent the scatter bands at 3 and 1/3.

**5.1 Results of steel samples**

Table 1 presents the performance of several machine learning models for the three different measures (stress, strain, and energy) based on different metrics. The aim is to foster a comprehensive understanding of how different algorithms perform under varying conditions, thereby providing insights into which algorithm and measure pair exhibits the best performance in terms of predictability and precision. It is expected that this detailed presentation of results will contribute significantly to the ongoing discourse in fatigue lifetime prediction using machine learning methods.

Table 1. Evaluation of ML Algorithms for fatigue lifetime prediction with different measures: results for steel samples.

|  | Model | $R^2$ (enhanced) | MSE (enhanced) | Best $R^2$ (enhanced) | $R^2$ (baseline) | MSE (baseline) | Best $R^2$ (baseline) |
|---|---|---|---|---|---|---|---|
| Stress | GB | 0.90 | 9.29e+09 | 0.95 | 0.45 | 6.50e+10 | 0.97 |
|  | RF | 0.89 | 9.72e+09 | 0.94 | 0.51 | 6.14e+10 | 0.95 |
|  | RL | 0.75 | 2.24e+10 | 0.85 | -0.26 | 1.37e+11 | 0.78 |
|  | MLP | 0.86 | 1.22e+10 | 0.95 | 0.34 | 8.01e+10 | 0.94 |
|  | VR | 0.89 | 1.01e+10 | 0.94 | 0.47 | 6.72e+10 | 0.93 |
| Strain | GB | 0.89 | 8.16e+09 | 0.95 | 0.41 | 6.65e+10 | 0.96 |
|  | RF | 0.89 | 8.46e+09 | 0.95 | 0.50 | 5.94e+10 | 0.94 |
|  | RL | 0.73 | 2.11e+10 | 0.86 | -0.27 | 1.37e+11 | 0.78 |
|  | MLP | 0.85 | 1.20e+10 | 0.96 | 0.34 | 8.20e+10 | 0.89 |
|  | VR | 0.88 | 9.02e+09 | 0.94 | 0.47 | 6.56e+10 | 0.90 |
| Energy | GB | 0.89 | 6.80e+09 | 0.95 | 0.31 | 7.56e+10 | 0.97 |
|  | RF | 0.89 | 7.02e+09 | 0.96 | 0.36 | 7.53e+10 | 0.96 |
|  | RL | 0.68 | 2.08e+10 | 0.79 | -0.53 | 1.51e+11 | 0.63 |
|  | MLP | 0.83 | 1.09e+10 | 0.95 | 0.20 | 9.13e+10 | 0.85 |
|  | VR | 0.87 | 8.13e+09 | 0.94 | 0.32 | 7.95e+10 | 0.93 |

Based on Table 1, in examining the stress measure, GB excels on the enhanced dataset with the highest $R^2$ score (0.90), which represents a high degree of correlation between predicted and actual values. This model also produces the lowest MSE (9.29e+09), implying fewer prediction errors, and achieves the best $R^2$ score (0.95). Nevertheless, for the baseline dataset, RF outperforms the other models with the highest $R^2$ score (0.51) and the smallest MSE (6.14e+10). For the strain measure, both GB and RF models perform exceptionally well on the enhanced dataset with identical $R^2$ scores (0.89). On the baseline dataset, RF again performs best, with the highest $R^2$ score (0.50) and the smallest MSE (5.94e+10). Notably, GB also produces the smallest MSE (8.16e+09) and best $R^2$ score (0.95) on the enhanced dataset, reinforcing its robustness across different measures. When evaluating the energy measure, GB and RF models continue to dominate. Both models generate the highest $R^2$ scores (0.89) on the enhanced dataset, while RF, once again, takes the lead on the baseline dataset with the highest $R^2$ score (0.36) and smallest MSE (7.53e+10). The GB model, consistent with the previous measures, delivers the lowest MSE (6.80e+09) and highest Best $R^2$ score (0.95) for the enhanced dataset.

The RL model consistently underperforms across all measures and datasets, suggesting it may not be suitable for this specific task due to its simplicity, as it assumes a linear relationship between input features. In contrast, the GB and RF models, which are known for their ability to handle non-linearity in data, excel in this task. The MLP and VR models deliver moderate performance, with $R^2$ scores that are consistently lower than those of GB and RF, but higher than RL. This suggests these models' performances fall in between, implying they might benefit from further tuning or the addition of more relevant features.

After evaluating the performance of models using $R^2$ and MSE scores, RF is utilized to illustrate estimations for unseen data. In this context, a random selection of 10% of the data is made to represent the unseen portion. Fig. 3 illustrates the models' predictions for the dataset related to steel.

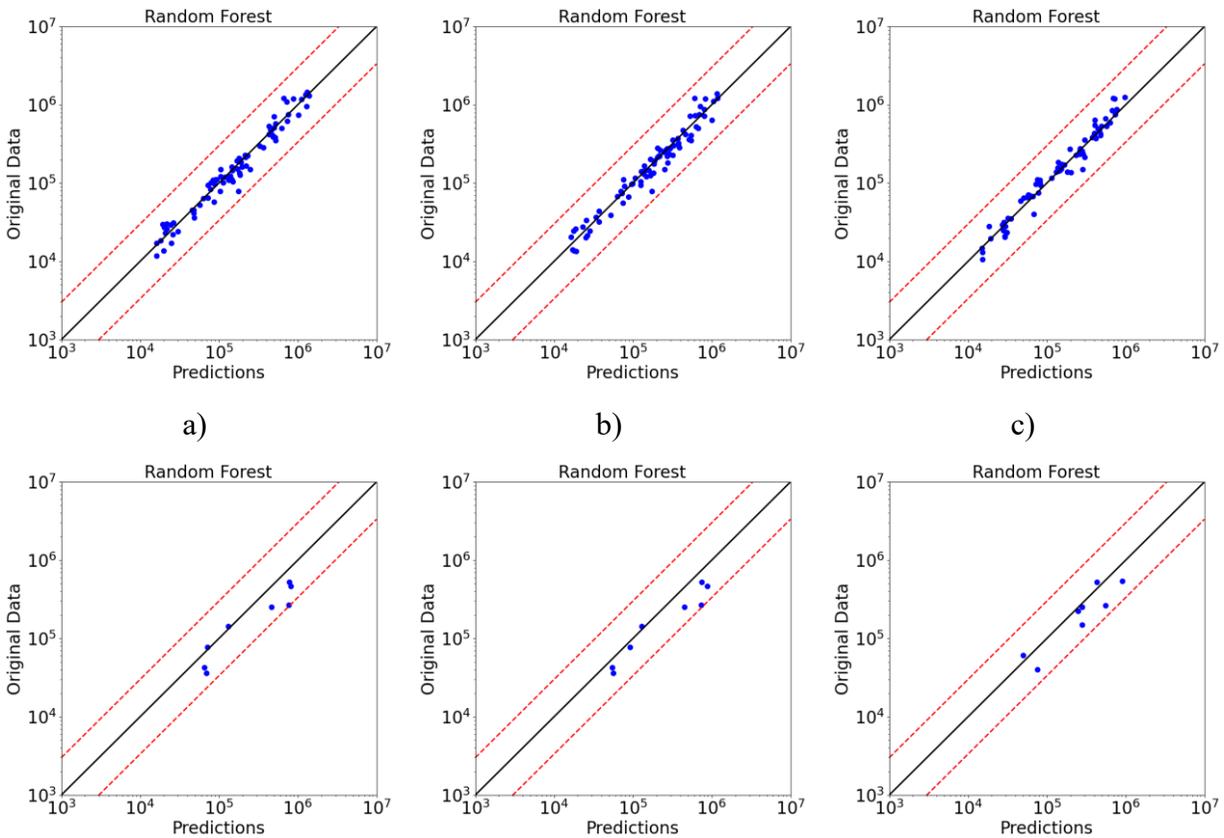

|   |   |   | d) |   | e) |   | f) |
|---|---|---|---|---|---|---|---|

Fig. 3. Comparison of experimental number of cycles to failure and predictions for steel samples using RF. The figure shows randomly selected samples with 10% of the dataset as unseen data. The analysis includes stress (a), strain (b), and energy (c) as measures for the enhanced dataset, and stress (d), strain (e), and energy (f) as measures for the baseline dataset.

Fig. 3 supports the findings presented in Table 1. It confirms that, for the enhanced data, energy is identified as the most effective measure. Conversely, for the baseline data, strain (followed closely by stress) emerges as the preferred measure. It is worth noting that, despite variations in notch geometries, loading conditions, and specimen thicknesses, all measures consistently yield promising results for predicting fatigue lifetime. In Section 5.5.4, a comprehensive discussion will be provided to explain why the enhanced data exhibits higher accuracy compared to the baseline data.

## 5.2 Results of PLA samples

Regarding the dataset related to PLA samples, Table 2 provides a thorough comparison of $R^2$ and MSE scores for different measures and algorithms for the analysis of PLA data. By presenting the comprehensive set of performance metrics, Table 2 facilitates a deeper comprehension of the respective strengths and weaknesses of each model within the specified measures.

Table 2. Evaluation of ML Algorithms for fatigue lifetime prediction with different measures: results for PLA samples.

|        | Model | $R^2$ (enhanced) | MSE (enhanced) | Best $R^2$ (enhanced) | $R^2$ (baseline) | MSE (baseline) | Best $R^2$ (baseline) |
|--------|-------|------|----------|----------|------|----------|------|
|        | GB    | 0.86 | 4.08e+10 | 0.97     | 0.57 | 8.78e+10 | 1.00 |
|        | RF    | 0.87 | 3.60e+10 | 0.97     | 0.61 | 8.13e+10 | 0.97 |
| Stress | RL    | 0.42 | 1.65e+11 | 0.59     | -0.05| 1.57e+11 | 0.71 |
|        | MLP   | 0.82 | 5.21e+10 | 0.97     | 0.45 | 9.66e+10 | 0.92 |
|        | VR    | 0.84 | 4.77e+10 | 0.96     | 0.58 | 8.29e+10 | 0.94 |
|        | GB    | 0.86 | 3.83e+10 | 0.98     | 0.53 | 9.36e+10 | 1.00 |
| Strain | RF    | 0.88 | 3.19e+10 | 0.98     | 0.59 | 8.39e+10 | 0.99 |
|        | RL    | 0.42 | 1.59e+11 | 0.58     | -0.05| 1.57e+11 | 0.71 |

|  | MLP | 0.82 | 5.00e+10 | 0.97 | 0.48 | 9.36e+10 | 0.93 |
|  | VR | 0.84 | 4.41e+10 | 0.94 | 0.57 | 8.40e+10 | 0.92 |
|  | GB | 0.78 | 6.20e+10 | 0.97 | 0.38 | 1.10e+11 | 0.99 |
|  | RF | 0.82 | 5.18e+10 | 0.96 | 0.48 | 9.90e+10 | 0.98 |
| Energy | RL | 0.40 | 1.62e+11 | 0.60 | -0.05 | 1.57e+11 | 0.71 |
|  | MLP | 0.78 | 6.19e+10 | 0.95 | -1.30 | 3.19e+11 | 0.86 |
|  | VR | 0.80 | 5.97e+10 | 0.94 | 0.50 | 9.33e+10 | 0.91 |

Examining the stress measure, RF outperforms other models on the enhanced dataset with the highest $R^2$ score (0.87), and the lowest MSE (3.60e+10). Interestingly, both GB and RF achieve the same Best $R^2$ score (0.97). On the baseline dataset, RF performs well again with the highest $R^2$ score (0.61) and the smallest MSE (8.13e+10). GB, however, reaches the maximum Best $R^2$ score of 1.00. Moving on to the strain measure, RF consistently delivers excellent results. It leads on the enhanced dataset with an $R^2$ score of 0.88 and the smallest MSE (3.19e+10). For the baseline dataset, RF also exhibits superior performance with the highest $R^2$ score (0.59) and the smallest MSE (8.39e+10). Both GB and RF achieve nearly perfect Best $R^2$ scores on both datasets, marking them as standout performers. When it comes to energy as a measure, RF comes out on top for the enhanced dataset with the highest $R^2$ score (0.82) and the lowest MSE (5.18e+10).

On the baseline dataset, VR achieves the highest $R^2$ score (0.50), while RF retains the smallest MSE (9.90e+10). The GB model, on the other hand, reaches the highest Best $R^2$ score for both datasets, reiterating its capability to achieve flawless predictions under certain conditions. The RL model shows a consistent underperformance for all measures and datasets, which indicates a possible inability to capture complex relationships in the data. Conversely, GB and RF models, known for handling non-linear data patterns effectively, prove their worth in this context. The MLP and VR models demonstrate average performance, with $R^2$ scores and MSEs generally lower than GB and RF, but higher than RL. This might indicate that these models might have room for optimization or might perform better with additional or different input features.

Fig. 4 visually demonstrates the results obtained from the analysis of both the enhanced and baseline data using the RF algorithm, for different measures of stress, strain, and energy for the dataset relating to PLA samples.

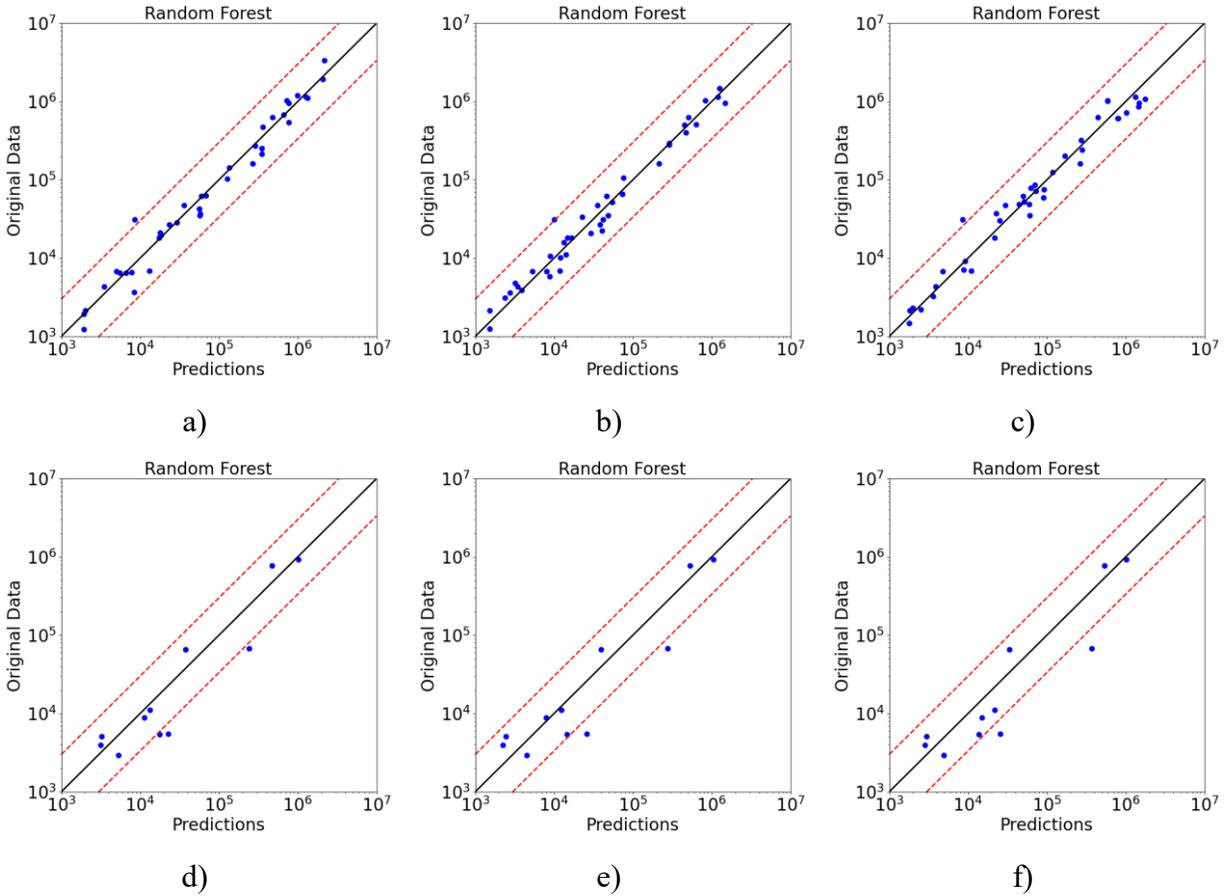

Fig. 4. Comparison of experimental number of cycles to failure and predictions for PLA samples using RF. The figure shows randomly selected samples with 10% of the dataset as unseen data. The analysis includes stress (a), strain (b), and energy (c) as measures for the enhanced dataset, and stress (d), strain (e), and energy (f) as measures for the baseline dataset.

Referring to Fig. 4, it can be observed that for the enhanced data, strain emerges as the most effective measure for predicting failure. Although stress yields similar results, the energy measure manifests slightly higher errors. Nevertheless, for all measures, the majority of the data falls within the scatter band of 3 and 1/3. This pattern indicates the robustness and validity of our models across different notch geometries and a wide spectrum of fatigue lives. For the baseline data, the results follow a similar trend, but in this case, stress proves to be the most effective measure. Overall, the outcome of this analysis is particularly encouraging as it underscores the models' ability to manage the high variance inherent in the baseline data, even when the effect of the manufacturing angle is not considered in our analysis. These findings demonstrate the potential and versatility of the framework in estimating fatigue lives.

## 5.3 Results of laminated composites samples

Here, a comprehensive evaluation of multiple predictive algorithms for various measures is provided, each applied to a dataset pertaining to composite materials. Table 3 exhibits the $R^2$ and MSE scores attained by each model for the different measures and across both the enhanced and baseline datasets. Through this comprehensive examination, the potential strengths and limitations of each algorithm when dealing with advanced composite materials under fatigue loading are demonstrated.

Table 3. Evaluation of ML Algorithms for fatigue lifetime prediction with different measures: results for laminated composites samples.

|  | Model | $R^2$ (enhanced) | MSE (enhanced) | Best $R^2$ (enhanced) | $R^2$ (baseline) | MSE (baseline) | Best $R^2$ (baseline) |
|---|---|---|---|---|---|---|---|
| Stress | GB | 0.93 | 2.51e+09 | 0.98 | 0.03 | 2.68e+10 | 0.98 |
|  | RF | 0.93 | 2.35e+09 | 0.98 | 0.27 | 2.12e+10 | 0.99 |
|  | RL | 0.72 | 9.84e+09 | 0.84 | -92.24 | 2.53e+11 | 0.96 |
|  | MLP | 0.84 | 5.95e+09 | 0.98 | -0.91 | 3.53e+10 | 0.99 |
|  | VR | 0.92 | 3.01e+09 | 0.97 | -9.96 | 4.77e+10 | 0.99 |
| Strain | GB | 0.90 | 3.27e+09 | 0.97 | 0.21 | 2.15e+10 | 0.99 |
|  | RF | 0.91 | 2.96e+09 | 0.98 | 0.35 | 2.03e+10 | 1.00 |
|  | RL | 0.73 | 9.17e+09 | 0.85 | -92.24 | 2.53e+11 | 0.96 |
|  | MLP | 0.81 | 6.46e+09 | 0.97 | -0.31 | 3.10e+10 | 0.96 |
|  | VR | 0.90 | 3.43e+09 | 0.96 | -10.22 | 4.63e+10 | 0.99 |
| Energy | GB | 0.92 | 2.37e+09 | 0.98 | -0.45 | 3.14e+10 | 0.97 |
|  | RF | 0.93 | 2.11e+09 | 0.98 | -0.91 | 2.97e+10 | 0.97 |
|  | RL | 0.74 | 8.19e+09 | 0.87 | -92.24 | 2.53e+11 | 0.96 |
|  | MLP | 0.67 | 1.07e+10 | 0.94 | -2.33 | 4.04e+10 | 0.98 |
|  | VR | 0.92 | 2.55e+09 | 0.98 | -11.42 | 5.40e+10 | 0.99 |

Looking at the stress measure, both GB and RF models exhibit superior performance on the enhanced dataset, achieving an identical $R^2$ score of 0.93, with RF registering a slightly lower MSE (2.35e+09). Moreover, both models attain an identical high Best $R^2$ score of 0.98. As for the baseline dataset, RF once again outshines the other models with an $R^2$ score of 0.27 and the smallest MSE (2.12e+10). RF also achieves the highest Best $R^2$ score (0.99) on the baseline dataset. Moving to the strain measure, RF maintains its strong performance. It outperforms the other models on the enhanced dataset with an $R^2$ score of 0.91 and the smallest MSE (2.96e+09). The GB and RF models both achieve nearly perfect Best $R^2$ scores, highlighting their superior predictive power. On the baseline dataset, RF once again leads with the highest $R^2$ score (0.35), the lowest MSE (2.03e+10), and an ideal Best $R^2$ score of 1.00. In the case of the energy measure, RF again delivers the highest $R^2$ score (0.93) and lowest MSE (2.11e+09) for the enhanced dataset, showing its consistent performance across different measures. On the baseline dataset, however, none of the models achieve a positive $R^2$ score, pointing to a potential difficulty in modeling the baseline energy measure accurately.

Overall, RL shows the poorest performance across all measures and datasets, having the lowest $R^2$ scores and highest MSEs. Conversely, the GB and RF models showcase their robust performance, especially RF, which leads in many categories. The MLP and VR models fall in between, not quite reaching the performance of the GB and RF models, but surpassing RL.

In order to further examine the models' performance, the next step involves estimating RF performance on a randomly selected 10% of the data, which was considered as unseen data, across different measures for both the enhanced and baseline datasets. The predictions of these models for the dataset associated with long carbon fiber composite materials are presented in Fig. 5.

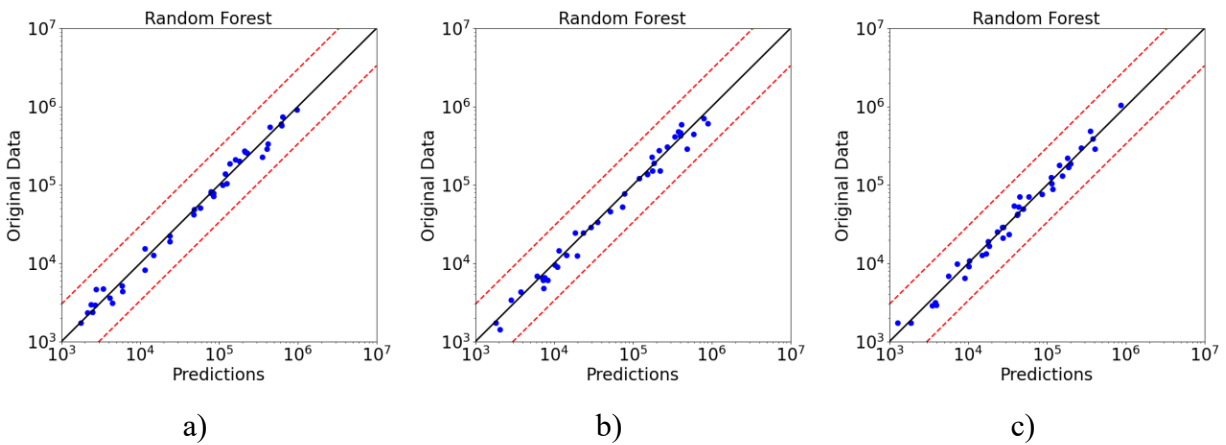

a)            b)            c)

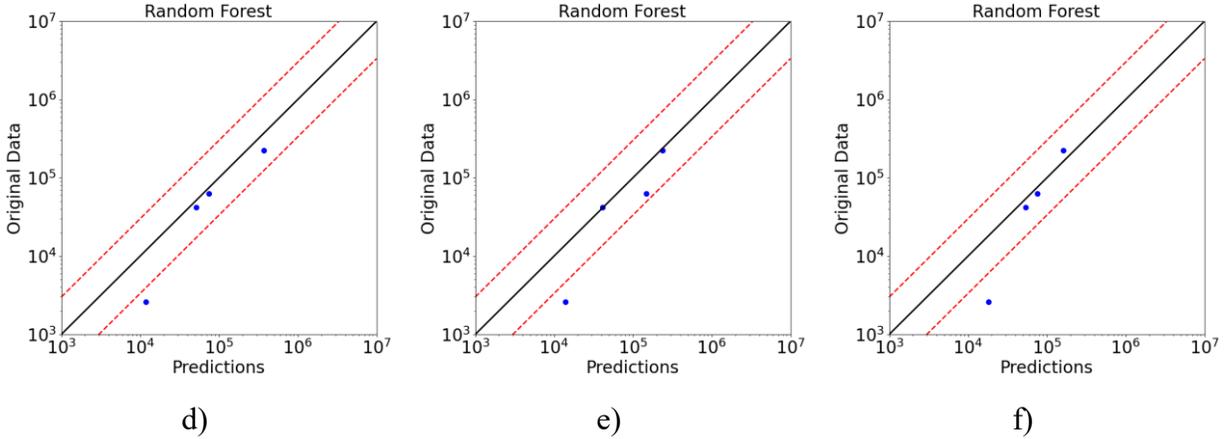

Fig. 5. Comparison of experimental number of cycles to failure and predictions for composites samples using RF. The figure shows randomly selected samples with 10% of the dataset as unseen data. The analysis includes stress (a), strain (b), and energy (c) as measures for the enhanced dataset, and stress (d), strain (e), and energy (f) as measures for the baseline dataset.

Observing Fig. 5, it becomes evident that for the unseen data, all of the measures deliver exceptional performance when applied to the enhanced dataset. All predictions fall within the scatter band of 3 and 1/3, indicating a high level of accuracy. However, the accuracy of predictions for the baseline dataset seems to falter, a conclusion that aligns with the results presented in Table 3. This discrepancy could be attributed to the small size of the dataset, which contains a total of just 33 data points. Furthermore, there are only four geometries in this dataset. Despite this, given the inherent high scatter observed in the analysis of composite materials, the obtained results can still be deemed satisfactory.

After a comprehensive examination of the provided tables, several salient points and patterns emerge that warrant discussion. First, it is evident that the ensemble models, GB and RF, consistently outperform the other models across all three materials. This upholds the principle of ensemble learning, which leverages the collective knowledge of numerous 'weak' learners to generate robust and reliable predictions. Second, there is a noticeable underperformance of the LR model compared to its counterparts. The complexities and potential non-linear patterns inherent in data may be a significant challenge for this model, leading to its diminished performance. This observation underscores the need for employing more sophisticated models capable of handling the inherent intricacies of such data. Third, an interesting trend is the performance improvement all models exhibit when applied to the enhanced dataset compared to the baseline dataset. This aligns with a common machine learning principle that a larger volume of high-quality data often leads to enhanced predictive accuracy. This observation may also suggest that the enhanced data could be inherently superior in quality (this behavior will be further discussed in detail later in this section). Fourth, the performance of the MLP and VR models varies across different scenarios. While they perform reasonably well in certain instances, they fall short in comparison to the ensemble methods. This inconsistency necessitates a more detailed investigation to pinpoint the conditions favorable for these models. Fifth and lastly, a notable pattern is the overall models' superior performance when utilizing stress as the measure, as opposed to strain or energy, across all three materials. This may indicate that the factors influencing stress are less complex or more linearly configured, thereby aligning favorably with the mathematical formulations used by these models. Additionally, the inherent structure of the stress-life diagrams used as inputs for the framework for all the measures may contribute to this trend.

## 5.4 A comparison with TCD

To gain a deeper understanding of the model's performance, it is essential to compare it with other existing methodologies. For this comparison, a well-established model grounded in fracture mechanics called the Theory of Critical Distances (TCD) [29,30] is utilized. The TCD employs stress distribution to predict the lifetime of notched components. It utilizes stress-life data from both plain and cracked (notched) samples to calibrate the inputs of the models and predict the fatigue lifetime of notched components. For brevity, only the last dataset (composite materials) is utilized, and the results of the TCD and the proposed framework considering only the RF algorithm are compared.

In most real-world situations, predicting the lifetime of an unseen notch geometry becomes necessary. The dataset related to composite materials comprises two distinct notched samples, as well as plain and cracked samples. For input characterization, TCD requires stress-life data from plain and cracked samples. However, in order to increase the number of data for employing RF for predictions, the data related to one notch geometry was designated as unseen data, while the other notch geometry, along with plain and cracked samples, was employed to train the model. This resulted in the code being executed twice, once for each notch geometry. The model was trained on the remaining portion of the data, augmented with enhanced data (100 data points for each type, except for the one being predicted). Fig. 6 displays the results of both models for these two notched geometries, along with their corresponding MSE values.

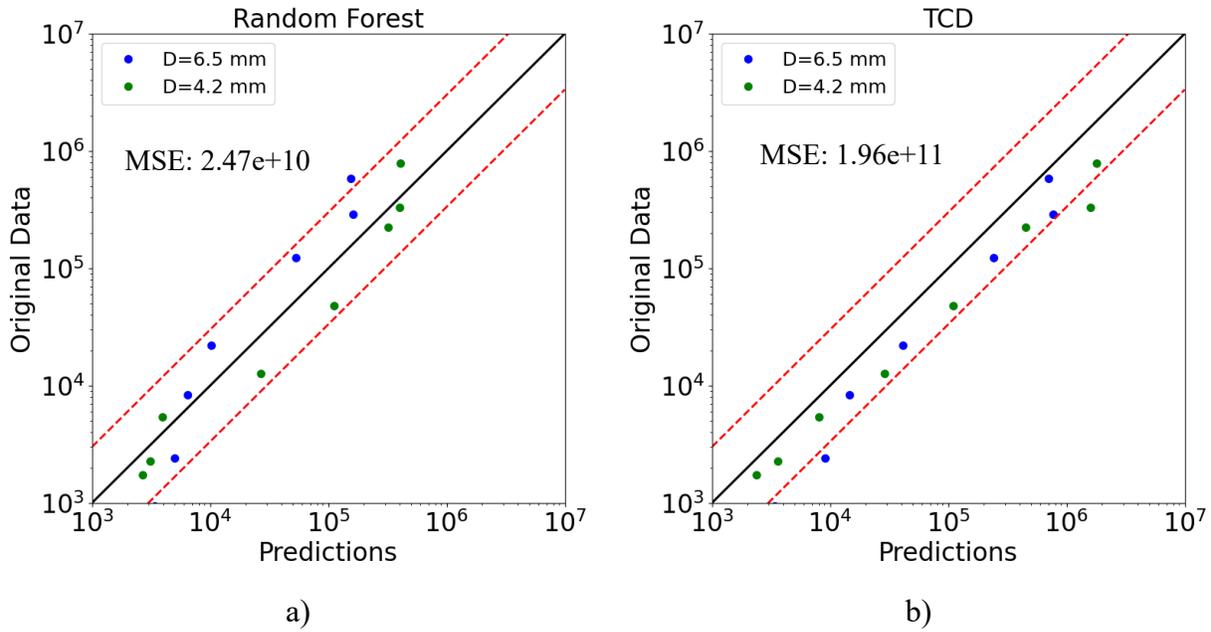

Fig. 6. Prediction of fatigue lifetime for notched components in composite materials based on: a) the proposed framework utilizing the RF algorithm, and b) TCD.

The comparative analysis presented in Fig. 6 showcases the superior predictive performance of the proposed framework, by employing the RF algorithm, as compared to the TCD. This advantage is underscored by a significant reduction in the Mean Squared Error (MSE) values for the proposed model (2.47e+10) as opposed to the TCD model (1.96e+11). Additionally, RF predictions exhibit a more balanced distribution around the solid line, representative of perfect predictions, thereby demonstrating the proposed model's proficiency in predicting the fatigue lifetime of notched components in composite materials. On the other hand, the TCD model's predictions tend to deviate below this line, landing in the unconservative region, thus suggesting potential underestimation of fatigue life and risking premature component failure.

## 5.5 Examination of key influential parameters

After demonstrating the effectiveness of the framework, it is necessary to study the influence of several key parameters, including the effect of gradient and number of data points, as well as path length. For this purpose, the data related to the composite materials are utilized, following the section related to the real-world case. However, for the sake of better comparison, only the results related to the notch geometry with *D*= 4.2 mm are presented. Note that in each subsection, only one parameter of interest is altered while the remaining parameters are kept consistent with those in the other analyses.

### 5.5.1 Effect of gradient of data

Four different types of data selection from the stress distribution are considered to analyze the effect of the gradient of data. Note that here, a higher gradient refers to a higher density of data close to the notch tip. In Fig. 7a, the stress distribution is shown as a solid black line, with different markers representing varying gradients. From Type I to Type IV, the number of points in closer proximity to the notch tip increases in order to better capture the stress concentration. In Type I, 10 points were chosen at equal distances.

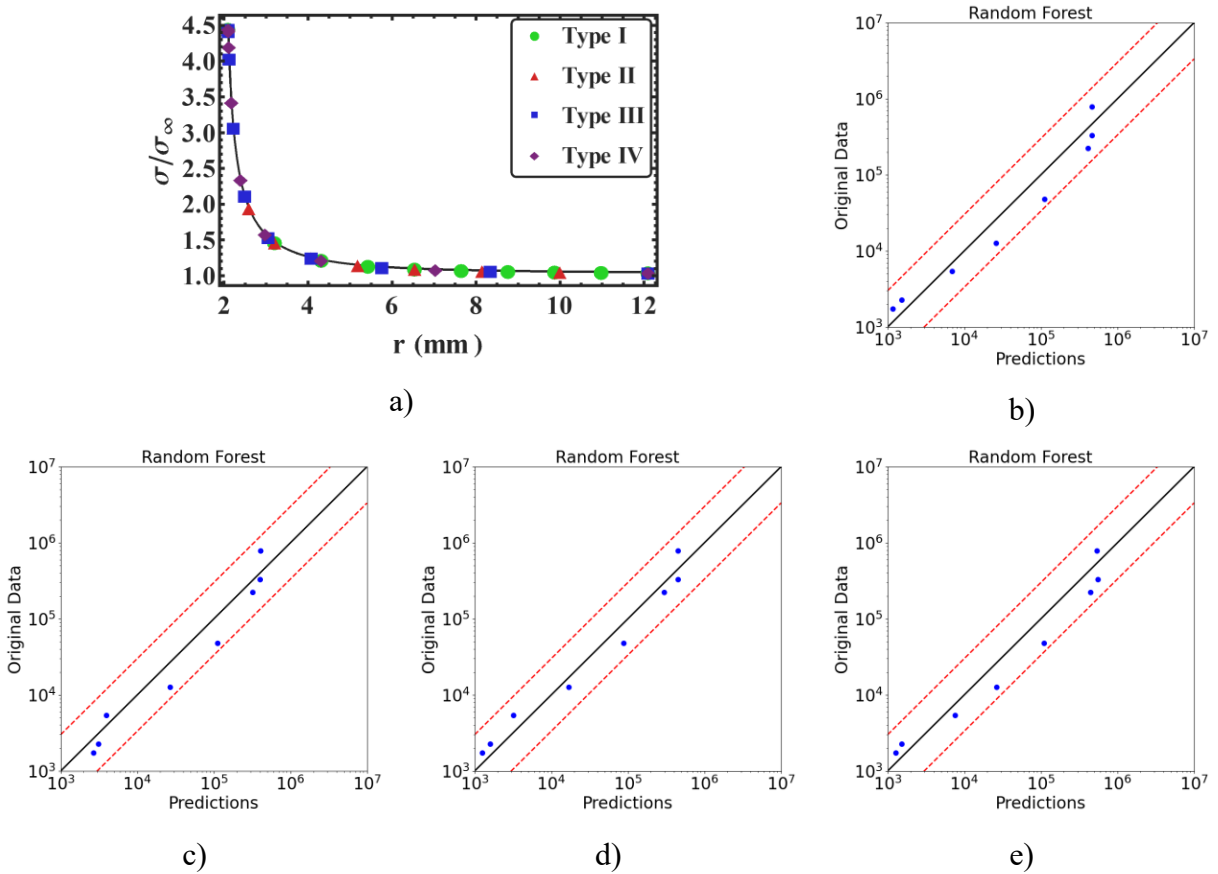

Fig. 7. a) Visualization of the stress distribution and data selection gradients. The solid black line represents the stress distribution, with varying data selection types represented by different points

along the line. b) results for type I, c) results for type II, d) results for type III, and e) results for type IV.

From the analysis presented in Fig. 7, it is clear that the model's performance remains remarkably consistent despite variations in data gradient. This illustrates the robustness of the model and underscores its stability and reliability. Regardless of the alteration in data density close to the notch tip—a transition from Type I to Type IV—the model effectively preserves its predictive power, offering accurate estimates of fatigue lives. As mentioned in Section 4, our analyses employed a gradient similar to Type II.

### 5.5.2 Effect of number of data points

The quantity of data points representing stress, strain, or energy distributions is also optional. As previously stated, 10 data points were considered for each case in the analyses. However, Fig. 8 depicts the impact of varying the quantity of data points on the predictive capability of the framework. It is noteworthy that, in all cases (except panel a)), the first and last data points of the path along the notch bisector line are used as inputs, and the number of points in between is different. As in the literature the peak stress at the notch tip is frequently used to estimate the fatigue life, panel a) represents this case.

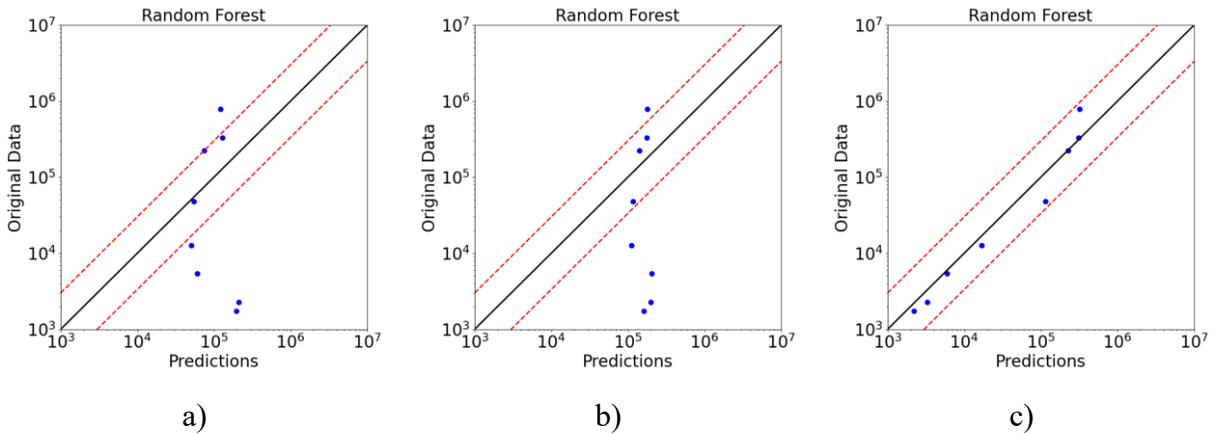

a)          b)          c)

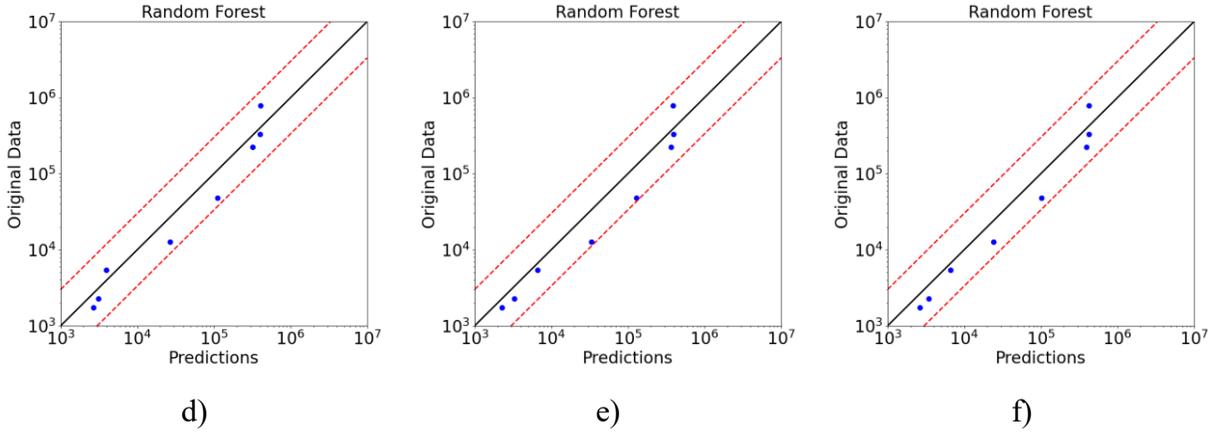

|       |       |       |
|-------|-------|-------|
| d)    | e)    | f)    |

Fig. 8. Performance analysis of the model for different number of data points: a) one, b) two, c) five, d) ten, e) fifteen, and f) twenty.

With the exception of cases a and b, the model's performance appears to remain largely stable, regardless of the number of data points, suggesting its effective capacity to capture and interpret essential data patterns. This characteristic not only confirms the model's reliability but also underlines its flexibility. Conversely, a comparison between cases a and b with the other scenarios underscores the importance of considering some intermediate data points between the first and last ones to better reflect the gradient of the field in the analysis.

### 5.5.3 Effect of path length

The selection of input data also involves determining the distance from the notch tip from which data should be selected. This decision inherently influences both the representation of the stress distribution and the consequent performance of the model. To meticulously investigate this parameter, data were collected from three distinct distances from the notch tip: 5 mm, 7.5 mm, and 10 mm. The outcomes of these considerations are depicted in Fig 9.

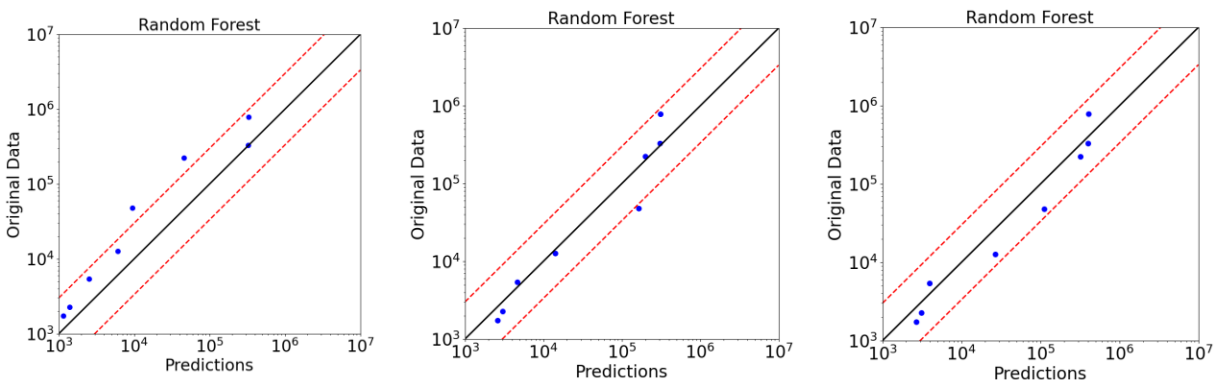

a)                b)                c)

Fig. 9: Performance analysis of the model for different distances from the notch tip (path lengths): a) 5 mm, b) 7.5 mm, c) 10 mm.

Fig. 9 offers an enlightening representation of the model's performance in relation to variations in the path length from the notch tip. As can be inferred from the figure, there is a positive correlation between an increase in path length and an enhancement in model performance. This trend can be scientifically rationalized. In the context of stress distribution, when the path length is sufficiently long, it allows the stress distribution to converge towards a fixed value which is for example the applied stress amplitude for sufficiently large ligament. This convergence is vital because it enables the model to capture the transition from the localized stress concentration around the notch to the global stress state. It provides a comprehensive view of the stress scenario, encompassing both the localized stress near the notch tip and the uniform stress distribution at a distance from it. This broad perspective enhances the model's predictive capability, facilitating a more comprehensive view of behavior under various loading conditions, thereby improving its performance.

### 5.5.4 Effect of number of generated data

This segment investigates how the quantity of data points in the enhanced dataset affects the model's performance. Similar to previous scenarios, the results pertain to the composite material, however, the unseen data were chosen randomly. The other parameters remain consistent with the primary analyses throughout the paper. Six different cases are considered, with the number of data points increased from the original dataset to 500 for each sample. The outcomes of this study are displayed in Fig. 10.

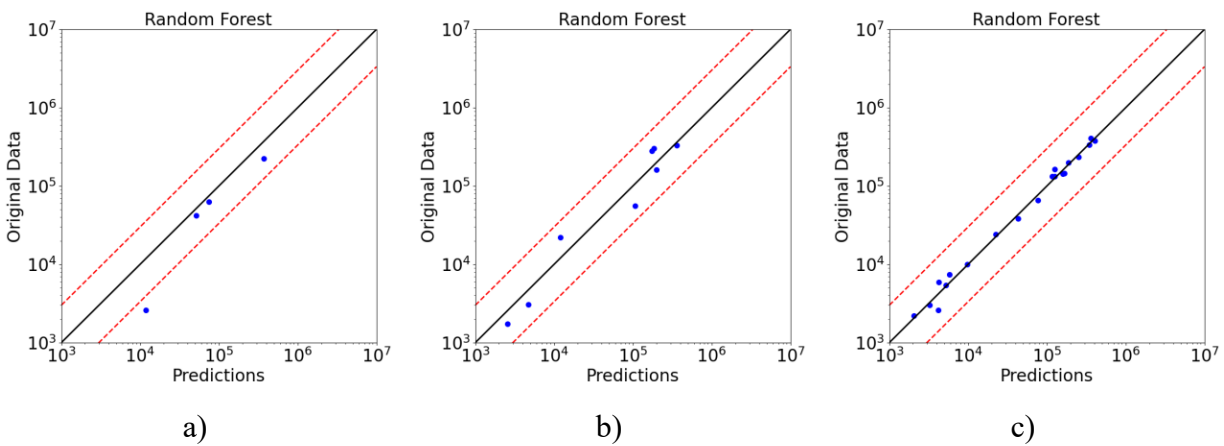

a)                b)                c)

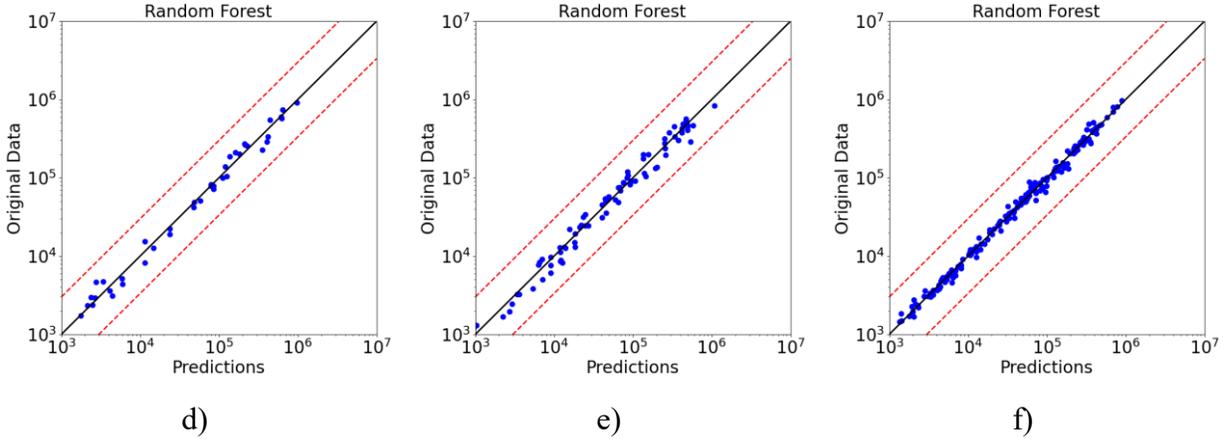

Fig. 10: Performance analysis of the model for different quantities of data points in the enhanced dataset: a) none, b) 20, c) 50, d) 100, e) 200, and f) 500.

The figure demonstrates a significant increase in the model's accuracy as data points are augmented, despite considerable noise in the generation of new data. This enhancement in performance of the proposed machine learning framework with increasing the number of data points in the enhanced datasets can be attributed to several factors:

1. Expansion of training examples: Machine learning algorithms, such as Random Forests, typically exhibit superior performance when provided with a larger number of examples to learn from. In the context of a dataset with limited size, the model may suffer from a shortage of sufficient information to decipher the governing patterns within the problem.

2. Reduction of overfitting: Properly increasing the quantity of data can bolster the model's robustness by mitigating overfitting. Overfitting refers to a scenario where the model excessively adapts to the noise in the training data (considering the high scatter in fatigue data), leading to suboptimal performance on new, unseen data. The inclusion of additional data (and consequently, more noise) in the dataset strengthens the model's resistance to noise, thereby reducing the likelihood of overfitting.

3. Bias-Variance trade-off: This principle states that an increase in bias leads to a decrease in variance, and vice-versa. By generating more data, we have subtly increased the model's bias, making it more generalized. At the same time, this decrease in variance lowers the model's susceptibility to overfitting and enhances its performance on unseen data.

**Conclusions**

A robust framework based on machine learning algorithms was proposed for the estimation of fatigue lifetime of notched components. To differentiate different notch geometries, the framework employs the gradient of mechanical properties, such as the stress field, strain field, and energy release rate, as a measure. Three distinct datasets were used for validation, covering a wide range of materials, such as steel, PLA, and carbon fiber composites, and notch geometries under different (uniaxial) loading conditions. All measures utilized in the framework contribute to a reasonably accurate estimation of fatigue lifetime, however, the stress measure proved to be the most effective. The study highlighted the effectiveness of machine learning models within the framework, specifically Gradient Boosting and Random Forest. Furthermore, a comparison with the TCD model highlighted the superior predictive performance of the proposed framework. The impact of some parameters on the performance of the framework was carefully analyzed, providing valuable insights into its robustness and stability under various conditions. The framework consistently maintained its predictive accuracy, regardless of changes in the data gradient or the total data points available. However, it was demonstrated that the path length needs to be sufficiently long to capture both local and nonlocal features. It was observed that enhancing the number of data points, in accordance with Basquin equation (with some added noise), improves the framework's performance by possibly providing more training examples, reducing overfitting, and facilitating a better bias-variance trade-off.

This research forms a foundation for future studies that could explore alternate machine learning algorithms and optimization techniques, with the aim of further enhancing the performance of the fatigue life prediction framework. From a mechanical standpoint, a potential improvement could be achieved by extending the model to account for multi-axial and variable loading conditions, broadening its scope and applicability.

**Acknowledgements**

The author acknowledges the funding from the European Union's Horizon 2020 research and innovation program under the Marie Skłodowska-Curie grant agreement No 861061 – NEWFRAC Project. The author would like to thank Dr. M. Manav (mmanav@ethz.ch) for his valuable assistance in reviewing this paper and providing insightful comments.